\documentstyle[aps,psfig]{revtex}
\newcommand{\be}{\begin{equation}}
\newcommand{\ee}{\end{equation}}
\newcommand{\ben}{\begin{eqnarray}}
\newcommand{\een}{\end{eqnarray}}

\begin{document}

\twocolumn[\hsize\textwidth\columnwidth\hsize\csname
@twocolumnfalse\endcsname

\title{Scalar fields: from domain walls to nanotubes and fulerenes}
\author{Dionisio Bazeia}
\address{Departamento de F\'\i sica, Universidade Federal da Para\'\i ba\\
Caixa Postal 5008, 58051-970 Jo\~ao Pessoa, Para\'\i ba, Brazil}

\maketitle

\begin{abstract}
In this work we review some features of topological defects in field theory
models for real scalar fields. We investigate topological defects in models
involving one and two or more real scalar fields. In models involving a
single field we examine two different subclasses of models, which support
one or more topological defects. In models involving two or more real scalar
fields, we explore the presence of defects that live inside topological
defects, and junctions and networks of defects. In the case of junctions
of defects we investigte structures that simulate nanotubes and fulerenes.
Our investigations may also be used to describe nonlinear properties
of polymers, Langmuir films and optical solitons in fibers.

\end{abstract}
\vskip2pc]

\section{Introduction}

Research on topological defects in Field Theory was iniciated almost three
decades ago, and some of the main investigations may be found for instance
in Refs.~{\cite{b1,b2,b3,b4,b5}}.
In the case of topological defects that appears in
models described by real scalar fields in $(1,1)$
space-time dimensions, they are usually named kinks,
and are classical static solutions of the equations of motion. Their
topological behavior is related to the asymptotic form of the field
configurations, which has to differ in both the positive and negative space
directions. To ensure that the classical solutions have finite energy, one
requires that the asymptotic behavior of the solutions is identified
with minima of the potential that defines the system under consideration,
so in general the potential has to include at least two distinct minima
for the system to support topological solutions.

We can investigate real scalar fields in $(3,1)$ space-time dimensions, and
now the topological solutions are named domain walls. These domain walls are
bidimensional structures that carry surface tension, which is identified with
the energy of the classical solutions that spring in $(1,1)$ space-time
dimensions. The domain wall structures are supposed to play a role in
applications to several different contexts, ranging from the low energy scale
of condensed matter \cite{b6,b7,b8} up to the high energy scale
required in the physics of elementary particles, fields and cosmology
\cite{b1,b2,b3,b4,b5}.

There are at least three classes of models that support
kinks or domain walls. In the first class one deals with a single
real scalar field, and the topological solutions are structureless.
Examples of this are the sine-Gordon and $\phi^4$ models. In the second
class of models one also deals with a single
real scalar field, but now the systems comprise at least two distinct
domain walls. An example of this is the double sine-Gordon model,
which has been investigated for instance in Refs.~{\cite{mb,dst80,cgm83}}.
In the third class of models we deal with systems defined by two real scalar
fields, and now one opens two new possibilities: domain walls
that admit internal structure \cite{wit85,lsh85,mke,mor,brs,97,98,mor98,bbb99},
and junctions of domain walls, which appear in models of two fields when
the potential contains non-colinear minima, as recently investigated for
instance in
Refs.{\cite{99a,99b,99c,00m,00a,00b,00v,00c,agm,h,00d,bv01,bb01,n,nns01}}.

There are other motivations to investigate
domain walls in models of field theory, one of them being related to
the fact that the low energy world volume dynamics of branes in string and
M theory may be described by standard models in field
theory \cite{s1,s2,s3}. Besides, one knows that field theory models
of scalar fields may also be used to investigate properties of quasi-linear
polymeric chains, as for instance in the applications of
Refs.~{\cite{96,99,00,01}}, to describe solitary waves in ferroelectric
crystals, the presence of twistons in polyethylene,
and solitons in Langmuir films.

In the present work, in Sec.~{\ref{sec:gen}} we review some known facts
about models described by real scalar fields.
The investigations follow in Secs.~{\ref{sec:1field}} and {\ref{sec:2fields}},
where we search for topological structures that generate kinks and walls,
in models involving a single real scalar field, and two or more fields,
respectively.

\section{General Considerations}
\label{sec:gen}

In this work we are interested in field theory models that describe real
scalar fields and support topological solutions of the
Bogomol'nyi-Prasad-Sommerfield (BPS) type \cite{b,ps}.
In the case of a single real scalar field $\phi$, we consider
the Lagrange density
\be
\label{1f}
{\cal L}=\frac12\partial_{\alpha}\phi\partial^{\alpha}\phi-V(\phi)
\ee
Here $V(\phi)$ is the potential, which identifies the particular model
under consideration. We write the potential in the form
$V(\phi)=(1/2)\,W^2_{\phi}$,
where $W=W(\phi)$ is a smooth function of the field
$\phi$, and $W_\phi=dW/d\phi$. In a supersymmetric theory $W$ is the
superpotential, and this is the way we name $W$ in this work.

The equation of motion for $\phi=\phi(x,t)$ has the general form
\be
\label{em2}
\frac{\partial^2\phi}{\partial t^2}-
\frac{\partial^2\phi}{\partial x^2}+\frac{dV}{d\phi}=0
\ee
and for static solutions we get
\be
\label{ems}
\frac{d^2\phi}{dx^2}=W_{\phi}W_{\phi\phi}
\ee
It was recently shown in Ref.~{\cite{bms01}} that this equation
of motion is equivalent to the first order equations
\be
\label{emf}
\frac{d\phi}{dx}=\pm W_{\phi}
\ee
if one is searching for solutions that obey the boundary conditions
$\lim_{x\to-\infty}\phi(x)={\bar\phi}_i$ and
$\lim_{x\to-\infty}(d\phi/dx)=0$, where ${\bar\phi}_i$ is one among the
several vacua $\{{\bar\phi}_1,{\bar\phi}_2,...\}$ of the system.
In this case the topological solutions are BPS (+) and anti-BPS (-)
solutions. Their energies get minimized to the value $t^{ij}=|\Delta W_{ij}|$,
where $\Delta W_{ij}=W_i-W_j$, with $W_i$ standing for $W({\bar\phi}_i)$.
The BPS and anti-BPS solutions are defined by two vacuum states belonging
to the set of minima that identify the several topological sectors of
the model.

In the case of two real scalar fields $\phi$ and $\chi$ the potential
is written in terms of the superpotential, in a way such that
$V(\phi,\chi)=(1/2)\,W^{2}_{\phi}+(1/2)\,W^{2}_{\chi}$.
The equations of motion for static fields are
\be
\frac{d^{2}\phi}{dx^{2}}=W_{\phi}W_{\phi\phi}+W_{\chi}W_{\chi\phi}
\ee
\be
\frac{d^{2}\chi}{dx^{2}}=W_{\phi}W_{\phi\chi}+W_{\chi}W_{\chi\chi}
\ee
which are solved by the first order equations
\be
\frac{d\phi}{dx}=\pm W_{\phi}\,\;\;\;\;\;\;\;\;
\frac{d\chi}{dx}=\pm W_{\chi}
\ee
Solutions to these first order equations are BPS (+) and anti-BPS (-)
states. They solve the equations of motion, and have energy minimized to
$t^{ij}=|\Delta W^{ij}|$ as in the case
of a single field; here, however,
$\Delta W^{ij}=W(\phi_i,\chi_i)-W(\phi_j,\chi_j)$, since now we need
a pair of numbers $(\phi_i,\chi_i)$ to represent each one of
the vacuum states in the system of two fields. In the plane
$(\phi,\chi)$ we may have minima that are non colinear, openning the
possibility for junctions of defects. In the case of two real scalar fields,
we can find a family of first order equations that are equivalent to the pair
of second order equations of motion, but this requires that
the superpotential is harmonic, obeying $W_{\phi\phi}+W_{\chi\chi}=0$
\cite{bms01}.

\section{Models involving a single real scalar field}
\label{sec:1field}

We now turn attention to kinks and domain walls. A well known example
is given by the $\phi^4$ model, defined by the potential
$V(\phi)=(1/2)\,(\phi^2-1)^2$.
Here we are using natural units, and dimensionless fields and coordinates.
In this model the domain wall can be represented by the solution
$\phi_s(x)=\pm\tanh(x)$. The above potential can be written with the
superpotential $W(\phi)=\phi-\phi^3/3$, and the domain wall is of the
BPS or anti-BPS type. The wall tension corresponding to the BPS wall
is $t_s=4/3$.

We can also find structureless domain walls in other models, for instance
in the $\phi^6$ model, which is described by the potential
$V(\phi)=(1/2)\,\phi^2\,(\phi^2-1)^2$.
Here we have
$W(\phi)=(1/2)\phi^2-(1/4)\phi^4$, and the wall configurations are also of the
BPS type, and are given by ${\bar\phi}^2_s=(1/2)[1\pm\tanh(x)]$.
The wall tension is now ${\bar t}_s=1/4$. This potential was investigated
for instance in Ref.~{\cite{loh79}}.

We can build another class of models where the domain walls
engender other features. The next class is yet described by a single field,
but the systems may now support two or more different walls. An interesting
example of this is the double sine-Gordon model, which is defined
by the potential
\be
\label{dsg}
V_r(\phi)=\frac1{r+1}\bigl[4\,r\,\cos(\phi)+\cos(2\phi)\bigr]
\ee
where $r$ is a parameter, real and positive. This potential is periodic,
with period $2\pi$; for simplicity in the following we consider the
interval $-2\pi<\phi<2\pi$. The value $r=1$ distinguishes two regions,
the region $r\in(0,1)$ where the potential contains four minima,
and the region $r\geq1$, where the potential contains two
minima. For $r\in(0,1)$ the system supports two distinct
wall configurations, the large wall and the small wall, which distinguish
the two different barrier the model comprises in this case. The limits
$r\to 0$ and $r\to\infty$ lead us back to the sine-Gordon model. The double
sine-Gordon model has been considered in several distinct applications,
as for instance in Ref.~{\cite{mb,dst80,cgm83}}, where one investigates
magnetic solitons in superfluid $^3$He, kink propagation in a model for
poling in polyvinylidine fluoride, and properties related to the two
different kinks that appear in such polymeric chain.

To expose new features of the double sine-Gordon model we rewrite
Eq.~(\ref{dsg}) in the form \cite{bil}
\be
\label{dsgw}
V_r(\phi)=\frac2{1+r}[\cos(\phi)+r\,]^2
\ee
where we have omitted an unimportant r-dependent constant.
The model can be described by the superpotential
\be
W(\phi)=\frac{2}{\sqrt{1+r}}[\sin(\phi)+r\,\phi]
\ee
For $r$ in the interval $r\in(0,1)$ the minima of the potential
are the singular points of the superpotential, $dW/d\phi=0$. They are
periodic, and for $-2\pi<\phi<2\pi$ there are four minima, at the points
${\bar\phi}=\pm\pi\pm\,\alpha(r)$,
where $\alpha(r)=\cos^{-1}(r)$. For $r\geq1$ the minima
are at ${\bar\phi}=\pm\pi$, in the interval $-2\pi<\phi<2\pi$. A closer
inspection shows that for $0<r<1$ the local maxima at $\pm\pi$
and the minima $\pm\pi\pm\alpha(r)$ degenerate to the minima
$\pm\pi$ for $r=1$, and remain there for $r>1$.
Thus, the parameter $r$ induces a transition in the
behavior of the double sine-Gordon model. The value $r=1$
is the critical value, since it is the point where the system
changes behavior: for $r\in(0,1)$ this model supports
minima that desapear for $r\geq1$. We illustrate the double sine-Gordon
model in Fig.~1, where we depict the potential of Eq.~(\ref{dsgw})
for $r=1/3,2/3$ and for $r=1$.

\begin{figure}
\centerline{\psfig{figure=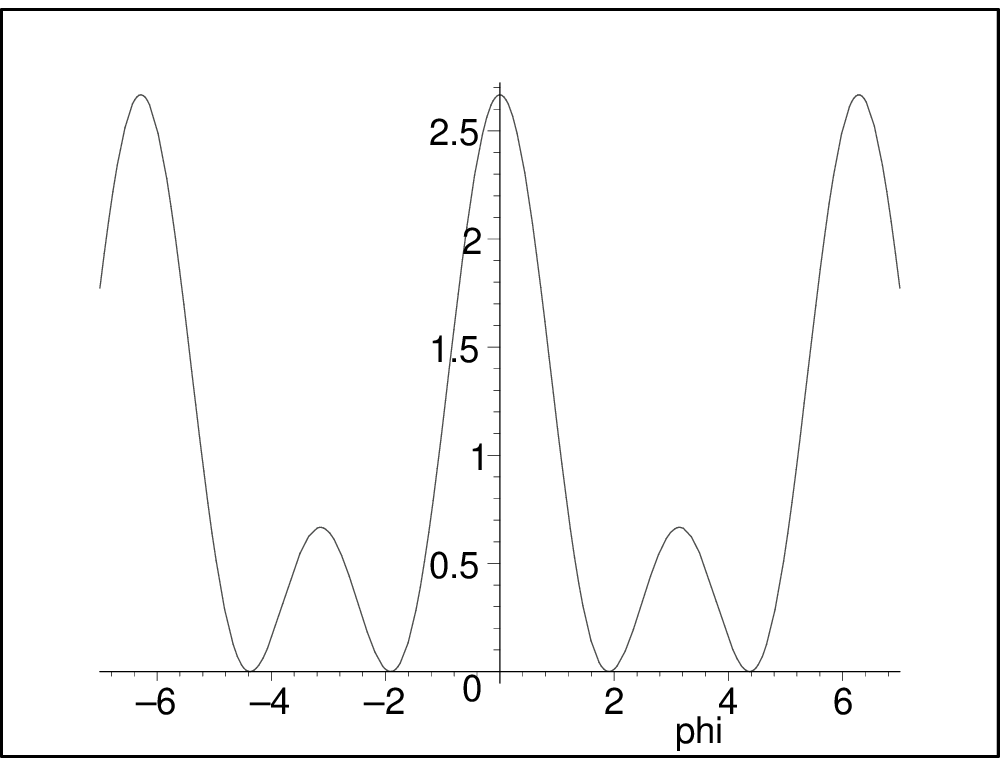,height=5.5cm}}
\centerline{\psfig{figure=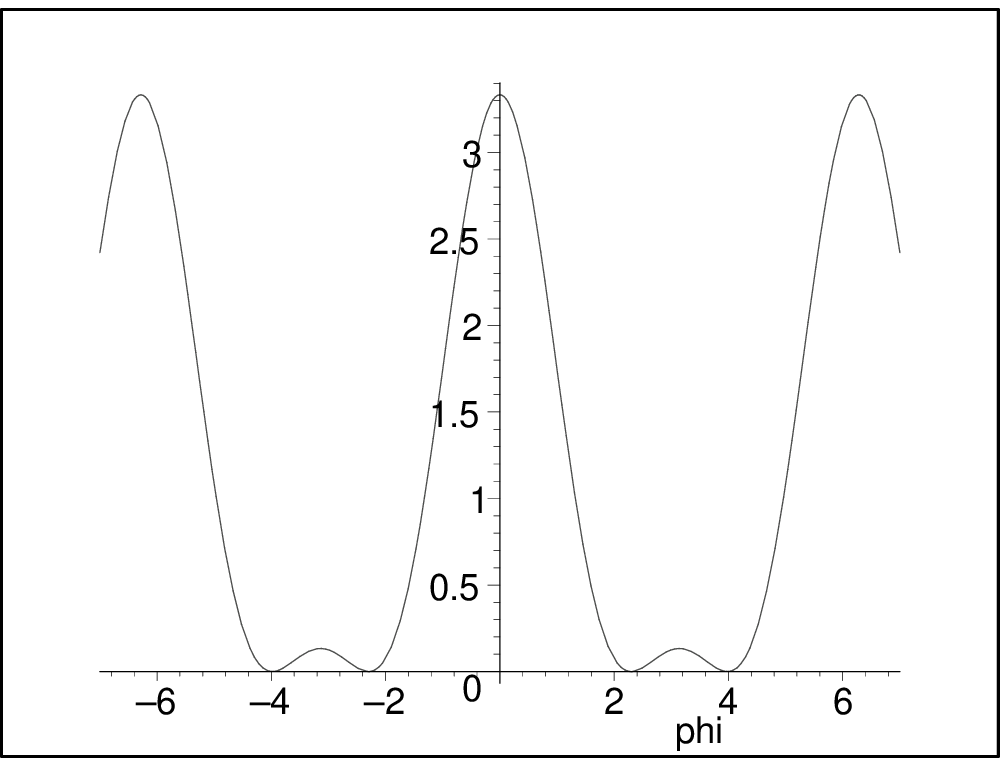,height=5.5cm}}
\centerline{\psfig{figure=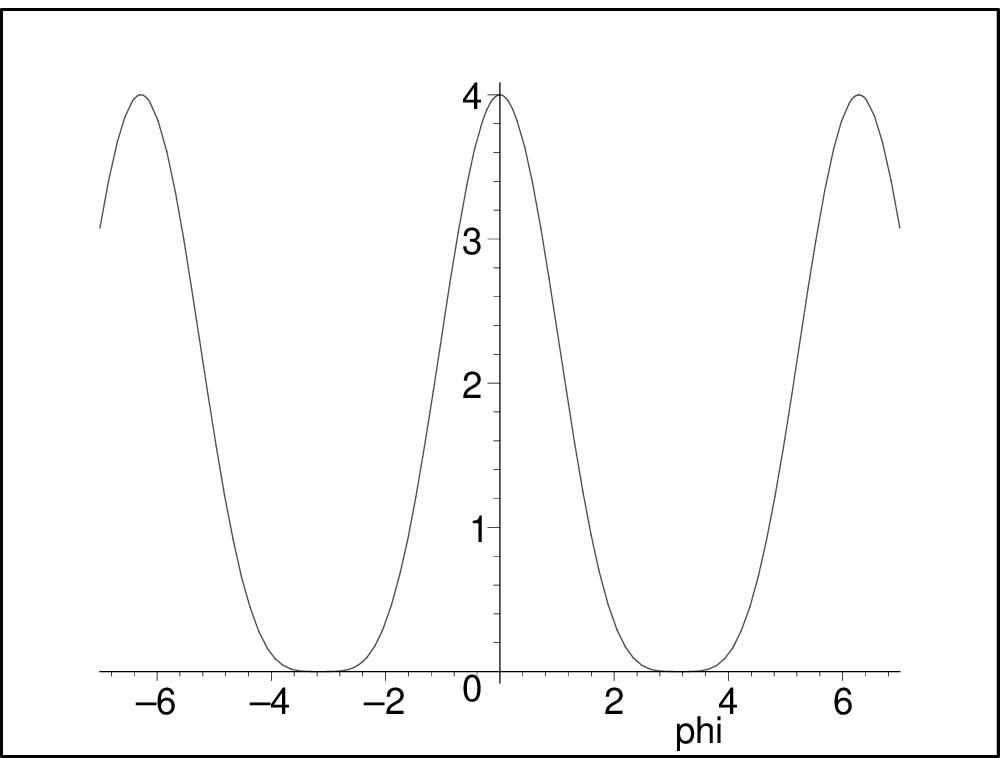,height=5.5cm}}
\vspace{0.3cm}
\caption{The double sine-Gordon potential, depicted for $r=1/3,2/3$ and
$1$ from above to below respectively, to illustrate how the behavior of the
model changes with $r$.}
\end{figure}

We get a better view of the phase transition in the double sine-Gordon
model by examining the order parameter ${\bar\phi}(r)$, which is given
by $ \pm\pi\pm\,\alpha(r)$ for $0<r\leq1$, so it goes continuously to
$\pm\pi$ for $r\geq1$. Also, the (squared) mass
of the field can be obtained via the relation
\be
V_r''({\bar\phi})= W^2_{{\bar\phi}{\bar\phi}}+
W_{{\bar\phi}}W_{{\bar\phi}{\bar\phi}{\bar\phi}}
\ee
where ${\bar\phi}$ is the corresponding minimum of the potential. For
$0<r\leq1$ we get $m^2(r)=4-4r$, and for $r\geq1$ we have
$m^2(r)=4(r-1)/(r+1)$. We plot ${\bar\phi}(r)$ and $m^2(r)$ in Fig.~2.
We see that $m(r)$ vanishes in the limit $r\to1$.
These results indicate that $r$ drives a second order phase transition,
a transition where the system goes from the case of two distinct phases
to another one, engendering a single phase.

\begin{figure}
\centerline{\psfig{figure=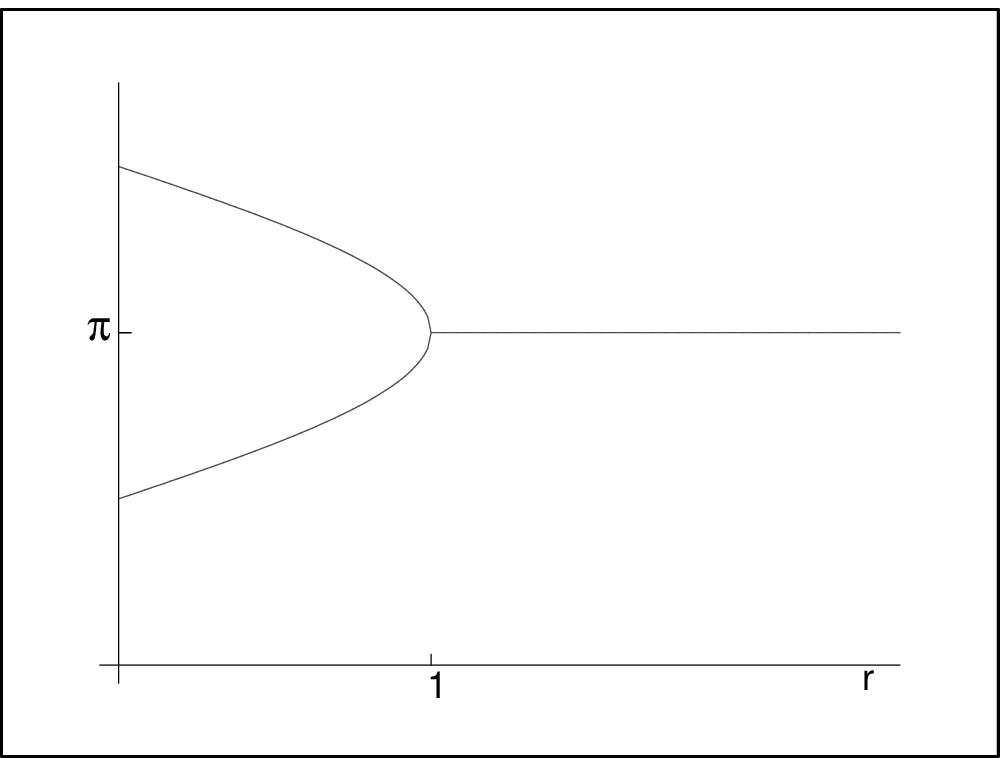,height=5.5cm}}
\centerline{\psfig{figure=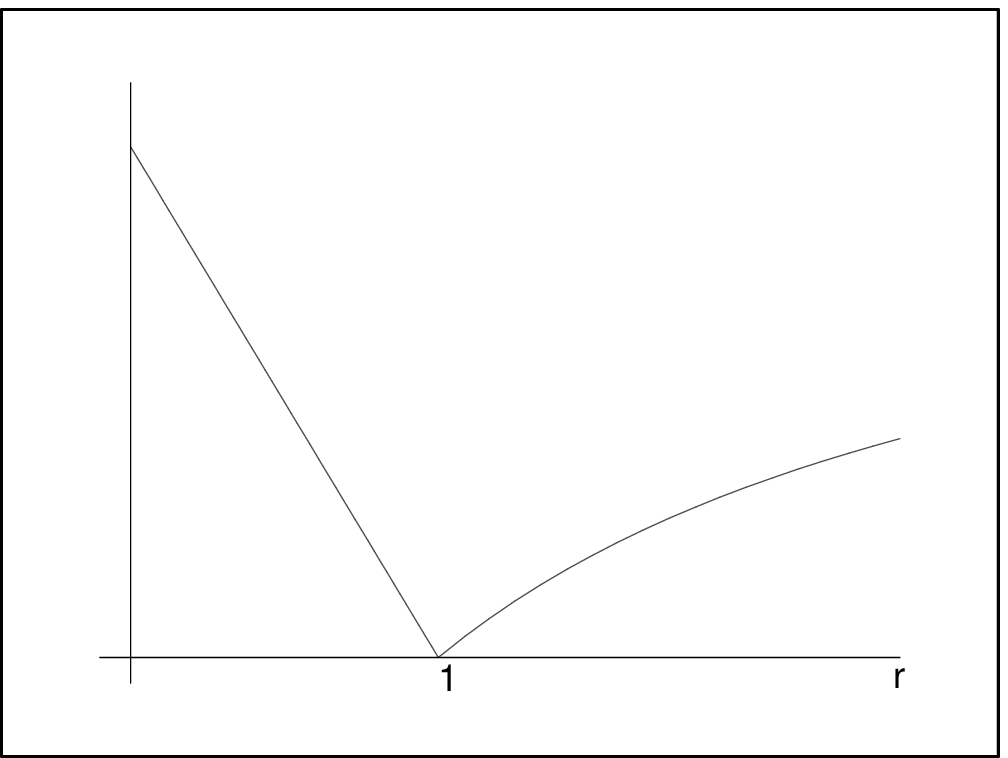,height=5.5cm}}
\vspace{0.3cm}
\caption{Plots of ${\bar{\phi}}(r)$ (above) and $m^2(r)$ (below) for the
double sine-Gordon potential, which illustrate how the behavior of the
model changes with $r$.}
\end{figure}

We consider $0<r\leq1$. The energies of the BPS solutions are given as follows.
For solutions connecting the minima $-\pi+\alpha(r)$ and $\pi-\alpha(r)$
the defect is large since it joins minima separated by a higher and wider
barrier. We have
\be
t^l_{dsG}=4\sqrt{1-r}+4r\frac{\pi-\alpha(r)}{\sqrt{1+r}}
\ee
In the case of the minima $\pi-\alpha(r)$ and $\pi+\alpha(r)$ the defect
is small and we get
\be
t^s_{dsG}=4\sqrt{1-r}-4r\frac{\alpha(r)}{\sqrt{1+r}}
\ee
We notice that $t^l_{dsG}=t^s_{dsG}+4\pi r/\sqrt{1+r}$, and that the limit
$r\to1$ sends $t^l_{dsG}\to 2\sqrt{2}\pi$ and $t^s_{dsG}\to0$, as expected.
For the BPS states we can write the solutions explicitly. For
instance, for solutions that connect the minima $-\pi+\alpha(r)$ and
$\pi-\alpha(r)$ we get large kink solutions, which are of the form
\be
\phi_{l}(x)=\pm 2\tan^{-1}\Biggl[\sqrt{\frac{1+r}{1-r}}\,
\tanh\left(\sqrt{1-r}\;x\right)\Biggr]
\ee
For solutions that connect the minima $\pi\pm\alpha(r)$ and the minima
$-\pi\pm\alpha(r)$ we get small kink solutions. They are given by
\be
\phi_{s}(x)=\pm\pi-2\tan^{-1}\Biggl[\sqrt{\frac{1-r}{1+r}}\,
\tanh\left(\sqrt{1-r}\,x\right)\Biggr]
\ee

The potential in Eq.~(\ref{dsgw}) in the limit $r\to0$ goes to
\be
V_0(\phi)=1+\cos(2\phi)
\ee
which leads us back to the sine-Gordon model. Thus, we can suppose
$r$ small and use $V_r(\phi)$ to explore the double sine-Gordon
model as a model controlled by a small parameter, in the vicinity
of the sine-Gordon model. This feature may be of some use for investigations
that follow the lines of Ref.~{\cite{abl01}}, and also in the case
concerning the presence of internal modes of solitary waves, which
seems to appear when one slightly modifies some integrable model --
see for instance Ref.~{\cite{k98}}.

\section{Models involving two or more real scalar fields}
\label{sec:2fields}

We now turn attention to the third class of models, which is described
by two real scalar fields. In this case the domain
walls may engender internal structure. This line of investigation follows
as in Refs.~{\cite{mke,mor,brs}} and we illustrate such possibility with
the system defined by the potential
\ben
\label{p2f}
V(\phi,\chi)&=&\frac12(\phi^1-1)^2+
\frac12r^2\left(\chi^2-\frac1r\right)^2+\nonumber\\
& &r(1+2r)\phi^2\chi^2
\een
where the parameter $r\neq0$ is real. This model was first investigated
in Ref.~\cite{95}, and can be used to modify the internal structure of
domain walls -- see Refs.~{\cite{brs,bnrt}}. This potential follows
from the superpotential
\be
\label{sp}
W(\phi,\chi)=\phi-\frac{1}{3}\phi^3-r\phi\chi^2
\ee
and the system supports two-field solutions. An explicit solution is
\ben
\phi(x)&=&\tanh(2rx)
\\
\chi(x)&=&a(r)\,{\rm sech}(2rx)
\een
with $a^2(r)=1/r-2$. This is a BPS solution, and now the parameter
$r$ is restricted to the interval $r\in(0,1/2)$. The limit
$r\to1/2$ lead us to the one-field solution
$\phi(x)=\tanh(x)$ and $\chi(x)=0$. As we know, all the BPS solutions
are linearly stable \cite{bs96}.

The two-field solutions obey
\be
\phi^2+\chi^2/a^2=1
\ee
which describes an elliptic arc connecting
the two minima $(\pm1,0)$ of the corresponding potential in the
$(\phi,\chi)$ plane. The one-field solutions represent standard
domain walls, while the two-field solutions may be seen as domain
walls having internal structure: the vector $(\phi,\chi)$ in configuration
space describes an straight line segment for the one-field solution, and
an elliptic arc for the two-field solution, resembling light in the
linearly and elliptically polarized cases, respectively. The same
solutions appear in condensed matter, in the anysotropic $XY$ model used
to describe ferromagnetic transition in magnetic systems, and there they
are named Ising and Bloch walls, respectively -- see for instance
Ref.~{\cite{b8}}, chapter 7.

We recall that domain walls may be seen as seeds \cite{b4,b5}
for the formation of non-topological structures.
This possibility appears in Refs.~{\cite{lee87,fgg88,mca95,mba96}},
where the discrete symmetry is changed to an approximate symmetry,
or in Ref.~{\cite{clo96}}, with the discrete symmetry biased so that
domains of distinct but degenerate vacua spring unequally. Domain walls
that appear in models involving two real scalar fields present features
that are not seen in the case of a single field. Thus, we are now
investigating other extensions of the above model, defined by the
superpotential of Eq.~(\ref{sp}). In the work in Ref.~{\cite{blw}}
we are investigating the possibility of changing the elliptic arc
to a more general orbit, that may find applications in condensed matter.
In another work \cite{blm} we are investigating other superpotentials,
non-polynomial, in an effort to extend the model studied in
\cite{choa,chob,baz99} to the case of two or more fields.
The model defined by the superpotential
of Eq.~(\ref{sp}) present other interesting features,
recently explored in Ref.~{\cite{sp02}}, which deserve further
investigations.

We now turn our attention to polynomial 
potentials that engenders the $Z_3$ symmetry, and that supports 
stable three-junctions that generate a regular hexagonal network of defects.
Investigations on the presence of junctions \cite{ato91} of defects in
supersymmetric models have been given in Refs.~{\cite{99a,99b,99c}}, and in
Ref.~{\cite{00a}} we have found the fourth-order polynomial potential 
\ben
V(\phi,\chi)&=&\lambda^2\phi^2\left(\phi^2-\frac{9}{4}\right)+ 
\lambda^2\chi^2\left(\chi^2-\frac{9}{4}\right)\nonumber  
\\ 
& &+\,2\,\lambda^2\,\phi^2\,\chi^2-\lambda^2\phi\,(\phi^2-3\,\chi^2)+ 
\frac{27}{8}\lambda^2  
\een
This potential does not allow for supersymmetric extensions.
The equations of motion for static configurations that follows in this case
are  
\ben  
\label{sm21}  
\frac{d^2\phi}{dx^2}&=&\lambda^2\phi\left(4\phi^2+ 
4\chi^2-3\,\phi-\frac{9}{2}\right)+3\lambda^2\chi^2  
\\  
\label{sm22}  
\frac{d^2\chi}{dx^2}&=&\lambda^2\chi\left(4\phi^2+4\chi^2+ 
6\,\phi-\frac{9}{2}\right)  
\een 
The potential has three degenerate minima, at the points $v_1=(3/2)\,(1,0)$ 
and $v_{2,3}=(3/4)\,(-1,\pm\sqrt{3})$. These minima form 
an equilateral triangle, invariant under the $Z_3$ symmetry. 
The distance between the minima is $(3/2)\sqrt{3}$. 
 
We can obtain the topological solutions explicitly. The easiest way 
to do this follows by first examining the sector that connects the vacua 
$v_2$ and $v_3$. This is so because 
in this case we set $\phi=-3/4$, searching for a strainght-line 
segment in the $(\phi,\chi)$ plane. This is compatible with the 
Eq.~(\ref{sm21}), and reduces the other Eq.~(\ref{sm22}) to the form 
\be 
\frac{d^2\chi}{dx^2}=\lambda^2\left(4\chi^3-\frac{27}{4}\chi\right) 
\ee 
This implies that the orbit connecting the vacua $v_2$ and $v_3$ 
is a straight line. It is such that, along the orbit the 
$\chi$ field feels the potential $\lambda^2\,[\chi^2-(27/16)]^2$. 
This shows that the model reduces to a model of a single field, 
and the solution satisfies the first-order equation 
\be 
\frac{d\chi}{dx}=\sqrt{2}\lambda\left(\chi^2-\frac{27}{16}\right) 
\ee 
The solution is 
\be 
\chi(x)=-\frac{3}{4}\sqrt{3}\tanh\left(\sqrt{\frac{27}{8}}\lambda\,x\right) 
\ee 
The other solutions can be obtained by rotations obeying the $Z_3$ symmetry 
of the model. 
 
The full set of solutions of the equations of motion are collected below.  
In the sector connecting the minima $v_2$ and $v_3$ they are  
\ben  
\label{sol11}  
\phi^{(\pm)}_{(2,3)}&=&-\frac{3}{4}  
\\  
\label{sol12}  
\chi^{(\pm)}_{(2,3)}&=&\pm\frac{3}{4}\sqrt{3}\, 
\tanh\left(\sqrt{\frac{27}{8}}\, \lambda\,x\right)  
\een  
In the sector connecting the minima $v_1$ and $v_2$ they are  
\ben  
\label{sol21}  
\phi^{(\pm)}_{(1,2)}&=&\frac{3}{8}  
\pm\frac{9}{8}\,\tanh\left(\sqrt{\frac{27}{8}}\,\lambda\,x\right)  
\\  
\label{sol22}  
\chi^{(\pm)}_{(1,2)}&=&\frac{3}{8}\sqrt{3}  
\mp\frac{3}{8}\sqrt{3}\,\tanh\left(\sqrt{\frac{27}{8}}\,\lambda\,x\right)  
\een  
In the sector connecting the minima $v_1$ and $v_3$ they are  
\ben  
\label{sol31}  
\phi^{(\pm)}_{(1,3)}&=&\frac{3}{8}  
\mp\frac{9}{8}\,\tanh\left(\sqrt{\frac{27}{8}}\,\lambda\,x\right)  
\\  
\label{sol32}  
\chi^{(\pm)}_{(1,3)}&=&-\frac{3}{8}\sqrt{3}  
\mp\frac{3}{8}\sqrt{3}\,\tanh\left(\sqrt{\frac{27}{8}}\,\lambda\,x\right)  
\een  
The label $(\pm)$ is used to identify kink and antikink. All the solutions  
have the same energy, $(9/4)\,\sqrt{27/8}\,|\lambda|$. 
 
We examine how the bosonic fields behave in the background 
of the classical solutions. We do this by considering fluctuations 
around the static solutions $\phi(x)$ and $\chi(x)$. 
We use the equations of motion to see that the fluctuations 
depend on the potential 
\be 
{\bf U}(x)={ {V_{\phi\phi}\,\,\, V_{\phi\chi} } 
\choose{ V_{\chi\phi}\,\,\, V_{\chi\chi} } } 
\ee 
Evidently, after obtaining the derivatives we substitute 
the fields by their classical static values $\phi(x)$ 
and $\chi(x)$. The model under consideration is defined 
by the potential (\ref{p}). In this case we use (\ref{sol11}) 
and (\ref{sol12}) to obtain two decoupled equations for the 
fluctuations. The potentials of the corresponding Schr\"odinger-like 
equations are 
\ben 
U_{11}(x)&=&\frac{27}{8}\lambda^2\Biggl[4-2\, 
{\rm sech}^2\left(\sqrt{\frac{27}{8}}\lambda\,x\right)\Biggr] 
\\ 
U_{22}(x)&=&\frac{27}{8}\lambda^2\Biggl[4-6\, 
{\rm sech}^2\left(\sqrt{\frac{27}{8}}\lambda\,x\right)\Biggr] 
\een 
The eigenvalues can be obtained explicitly: in the $\chi$ direction 
we get $w^{\chi}_0=0$ and $w^{\chi}_1=(9/2)\sqrt{\lambda^2/2}$, 
and in the $\phi$ direction we have $w^{\phi}_0=(9/2)\sqrt{\lambda^2/2}$. 
This shows that the pair $(\ref{sol11})$ and $(\ref{sol12})$ 
is stable, and by symmetry we get that all the three topological 
solutions are stable solutions. 
 
The classical solutions present the nice property of having energy 
evenly distributed in their gradient (g) and potential (p) 
portions. In terms of energy density they are 
\be 
{\rm g}(x)={\rm p}(x)=\frac{1}{4}\left(\frac{27}{8}\right)^2\,\lambda^2\,
{\rm sech}^{4}\left(\sqrt{\frac{27}{8}}\lambda\,x\right) 
\ee
To understand this feature we recall the calculation done 
explicitly in the sector with $\phi=-3/4$, constant. There the
model is shown to reduce to a model of a single field, a model
that supports BPS solutions. Within this context, the above 
solutions are very much like the non-BPS solutions that appear in 
supersymmetric systems \cite{00b}. We use this property and the 
topological current
\be
J^{\alpha}=\varepsilon^{\alpha\beta}\partial_{\beta}{\phi\choose\chi}
\ee
It obeys $\partial_{\alpha}J^{\alpha}=0$, and it is also a vector 
in the $(\phi,\chi)$ plane. For static configurations we have 
$J^{t}_{\alpha}\,J^{\alpha}=\rho^t\,\rho$, where $\rho=\rho(\phi,\chi)$ 
is the charge density. This charge density allows writing 
$\rho^t\rho=\varepsilon(x)$, 
where $\varepsilon(x)={\rm g}(x)+{\rm p}(x)$ is the (total) energy 
density of the solution. We use this result and the notation $ij$, 
to identify the sector connecting the vacua $(\phi_i,\chi_i)$ and
$(\phi_j,\chi_j)$, to show that for any two different sectors 
$ij$ and $jk$, $i,j,k=1,2,3$ we get that
\be
(\rho_{ij}+\rho_{jk})^t\,(\rho_{ij}+\rho_{jk})< 
\rho_{ij}^t\rho_{ij}+\rho_{jk}^t\rho_{jk}
\ee
This condition shows that the three-junction is a process
of fusion of defects that occurs exothermically, providing
stability of junctions in the present model. This result
is more general than the one in Ref.~{\cite{ato91}}, which
appears within the context of supersymmetry. Evidently, our
result also works for BPS and non-BPS solutions that appears in
supersymmetric models, with the property of having energy evenly
distributed in their gradient and potential portions \cite{00b}.
 
We notice that the orbits corresponding to the stable defect solutions  
form an equilateral triangle in the $(\phi,\chi)$ plane. This is so because 
the solutions are straight-line segments joining the three vacuum states  
in configuration space. They are degenerate in energy, and this allows
associating to each defect the same tension
\be
\label{ten}
t=\frac{9}{4}\,\sqrt{\frac{27}{8}}\,\,|\lambda|
\ee
This makes $t_{ij} < t_{jk}+t_{ki}, i,j,k=1,2,3$, and now the
inequality is strictly valid in this case, stabilizing the three-junction
that appears in this model when one enlarges the space-time to three
spatial dimensions.  
  
We consider the possibility of junctions in the plane, which may give  
rise to a planar network of defects. We work in $(2,1)$ space-time dimensions,
in the plane $(x,y)$. We identify the plane $(x,y)$ with the space of 
configurations, the plane $(\phi,\chi)$. We illustrate this situation by 
considering, for instance, the solutions  
we have already obtained. They are collected  
in Eqs.~(\ref{sol11})-(\ref{sol32}) in (1,1) dimensions.  
In the planar case they change to  
\ben  
\phi^{(\pm)}_{(2,3)}&=&-\frac{3}{4}  
\\  
\chi^{(\pm)}_{(2,3)}&=&\pm\frac{3}{4}\sqrt{3}  
\tanh\left(\sqrt{\frac{27}{8}}\,\lambda\,y\right)  
\een  
and  
\ben  
\phi^{(\pm)}_{(1,2)}&=&\frac{3}{8}\mp\frac{9}{8}  
\tanh\left(\frac{1}{2}\sqrt{\frac{27}{8}}\,\lambda\,(y+\sqrt{3}x)\right)  
\\  
\chi^{(\pm)}_{(1,2)}&=&\frac{3}{8}\sqrt{3}\pm\frac{3}{8}\sqrt{3}  
\tanh\left(\frac{1}{2}\sqrt{\frac{27}{8}}\,\lambda\,(y+\sqrt{3}x)\right)  
\een  
and  
\ben  
\phi^{(\pm)}_{(1,3)}&=&\frac{3}{8}\pm\frac{9}{8}  
\tanh\left(\frac{1}{2}\sqrt{\frac{27}{8}}\,\lambda\,(y-\sqrt{3}x)\right)  
\\  
\chi^{(\pm)}_{(1,3)}&=&-\frac{3}{8}\sqrt{3}\pm\frac{3}{8}\sqrt{3}  
\tanh\left(\frac{1}{2}\sqrt{\frac{27}{8}}\,\lambda\,(y-\sqrt{3}x)\right)
\een  
These planar defects are domain walls, and can be used to represent
the three-junction in the limit of thin walls.

The three-junction that appears in this $Z_3$-symmetric model allows  
building a network of defects, precisely in the form of a regular hexagonal  
network, as depicted in Fig.~3 in the thin wall approximation. In this
network the tension associated to the defect is the tipical value of the
energy in this tiling of the plane with a regular hexagonal network,
which seems to be the most efficient way of tiling the plane. As we
have shown, our model behaves standardly in $(3,1)$ dimensions.
It supports stable three-junctions that generate a stable regular
hexagonal network of defects.

\vspace{1.0cm} 
\begin{figure} 
\centerline{\psfig{figure=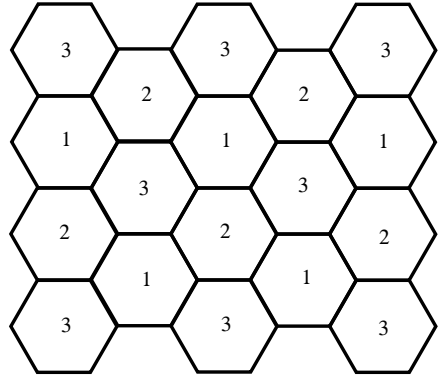,height=5.0cm}} 
\vspace{1.0cm} 
\caption{A regular hexagonal network of defects, formed by
three-junctions surrounded by domains representing the
vacua $v_1=1, v_2=2$, and $v_3=3$.} 
\end{figure}

In Ref.~{\cite{00a}} the idea of nesting a network of defects inside a
domain wall has been presented. A model that contains
the basic mechanisms behind this idea was introduced in Ref.~{\cite{00d}}.
It is described by three real scalar fields $\sigma$, $\phi$,
and $\chi$, and is defined by the (dimensionless) potential,
\ben
\label{p}
V(\sigma,\phi,\chi)&=&\frac{2}{3}\,\left(\sigma^2-\frac{9}{4}\right)^2+
\left(r\,\sigma^2-\frac{9}{4}\right)\,(\phi^2+\chi^2)
\nonumber\\
& &+(\phi^2+\chi^2)^2-\phi\,(\phi^2-3\,\chi^2)
\een
Here $r$ couples $\sigma$ to the pair of fields $(\phi,\chi)$. This potential
is polynomial, and contains up to the fourth order power in the fields.
Thus, it behaves standardly in $(3,1)$ space-time dimensions. Also, it
presents discrete $Z_2\times Z_3$ symmetry. We set $(\phi,\chi)\to(0,0)$,
to get the projection $V(\sigma,0,0)\to V(\sigma)=(2/3)\,(\sigma^2-9/4)^2$.
The projected potential presents $Z_2$ symmetry, and can be written with
the superpotential $W(\sigma)=(2\sqrt{3}/9)\sigma^3-(3\sqrt{3}/2)\sigma$,
in the form $V=(1/2)(dW/d\sigma)^2$. The reduced model supports the explicit
configurations $\sigma_h(z)=\pm\,(3/2)\,\tanh(\sqrt{3} z)$. The tension
of the host wall is $t_h=3\,\sqrt{3}=(3/2)\,m_h$,
where $m_h$ represents the mass of the elementary $\sigma$ meson.
Also, the width of the wall is such that $l_h\sim1/\sqrt{3}$.

The potentials projected inside $(\sigma\to0)$ and outside
$(\sigma\to\pm\,3/2)$ the host domain wall are $V_{in}(\phi,\chi)$
and $V_{out}(\phi,\chi)$. Inside the wall we have
\ben
V_{in}(\phi,\chi)&=&(\phi^2+\chi^2)^2-\phi\,(\phi^2-3\,\chi^2)\nonumber
\\
& &-\frac{9}{4}(\phi^2+\chi^2)+\frac{27}{8}
\een
This potential engenders the $Z_3$ symmetry, and there are three
global minima, at the points $v^{in}_1=(3/2)(1,0)$ and
$v^{in}_{2,3}=(3/4)(-1,\pm\sqrt{3})$, which define an
equilateral triangle. Outside the wall we get
\ben
V_{out}(\phi,\chi)&=&(\phi^2+\chi^2)^2-\phi\,
(\phi^2-3\,\chi^2)\nonumber\\
& &+\frac{9}{4}\,(r-1)\,(\phi^2+\chi^2)
\een
$V_{out}$ also engenders the $Z_3$ symmetry, but now the minima depend
on $r$. We can adjust $r$ such that $r>9/8$, which is the condition
for the fields $\phi$ and $\chi$ to develop no non-zero vacuum expectation
value outside the host domain wall, ensuring that the model supports no
domain defect outside the host domain wall. The restriction of considering
quartic potentials forbids the possibility of describing the $Z_3$ portion
of the model with the complex superpotential used in {\cite{99a}.

We investigate the masses of the elementary $\phi$ and $\chi$ mesons.
Inside the wall they degenerate to the single value $m_{in}=3\sqrt{3/2}$.
Outside the wall, for $r>9/8$ they also degenerate to a single value,
$m_{out}(r)=3\sqrt{(r-1)/2}$, which depends on $r$. We see that
$m_{out}(r=4)=m_{in}$. Also, $m_{out}(r)> m_{in}$ for $r$ bigger than 4,
and $m_{out}(r)< m_{in}$ for $r$ in the interval $(9/8,4)$.

We study linear stability of the classical solutions
$\sigma=\sigma_h(z)$ and $(\phi,\chi)=(0,0)$. The fields $\phi$ and
$\chi$ vanish classically, and their fluctuations $(\eta_n,\xi_n)$ decouple.
The procedure leads to two equations for the fluctuations,
that degenerate to the single Schr\"odinger-like equation
\be
-\,\frac{d^2\psi_n(z)}{dz^2}+\frac{9}{2}\,V(z)\;
\psi_n(z)=\,w_n^2\,\psi_n(z)
\ee
Here $V(z)=-1+r\,\tanh^2{\sqrt{3}z}$.
This equation is of the modified P\"oschl-Teller type, and can be
examined analytically. The lowest eigenvalue is
$w_0^2=(3/2)\sqrt{6\,r+1\,}-6$. There is instability for $r<5/2$, showing
that the host domain wall with $(\phi,\chi)=(0,0)$ is unstable and therefore
relax to lower energy configurations, with $(\phi,\chi)\neq(0,0)$ for
$r<5/2$. Inside the host domain wall the sigma field vanishes, and the model
is governed by the potential $V_{in}(\phi,\chi)$, which consequently may
allow the presence of non-trivial $(\phi,\chi)$ configurations. The host
domain wall entraps the system described by $V_{in}(\phi,\chi)$ for the
parameter $r$ in the interval $(9/8,5/2)$. In this interval we have
$m_{out}<m_{in}$, showing that it is not energetically favorable for the
elementary $\phi$ and $\chi$ mesons to live inside the wall for
$r\in(9/8,5/2)$. The model automatically suppress backreactions of the
$\phi$ and $\chi$ mesons into the defects that may appear inside
the host domain wall.

In Ref.~{\cite{00a}} the potential inside the wall was shown to admit
a network of domain walls, in the form of a hexagonal array of domain walls.
In the thin wall approximation the network may be
represented by the solutions
\ben 
\phi_{_1}&=&\frac{3}{8}+\frac{9}{8} 
\tanh\left(\frac{1}{2}\sqrt{\frac{27}{8}}\,(y+\sqrt{3}x)\right) 
\\ 
\chi_{_1}&=&\frac{3}{8}\sqrt{3}-\frac{3}{8}\sqrt{3} 
\tanh\left(\frac{1}{2}\sqrt{\frac{27}{8}}\,(y+\sqrt{3}x)\right) 
\een 
and by $(\phi_k,\chi_k)$, obtained by rotating the pair
$(\phi_1,\chi_1)$ by $2(k-1)\pi/3$, for $k=2,3$. We identify the
space $(\phi,\chi)$ with $(x,y)$, so rotations in $(\phi,\chi)$ also
rotates the plane $(x,y)$ accordingly. The energy or tension of the
individual defects in the network is given by, in the thin wall
approximation, $t_{n}=(27/8)\sqrt{3/2}=(9/8)\;m_{in}$. In the
nested network, the width of each defect obeys $l_n\sim\sqrt{8/27}$.
This shows that $l_h/l_n=3/2\sqrt{2}$, and so the host domain wall is
slightly thicker than the defects in the nested network. In the thin
wall approximation, the potential $V_{in}(\phi,\chi)$ allows the formation
of three-junctions as reactions that occur exothermically, and the nested
array of thin wall configurations is stable. In Fig.~4 we depict the
hexagonal network of defects inside the domain wall, in the thin wall
approximation; the dashed lines show equilateral triangles, that belong
to the dual lattice. Both the hexagonal network and the dual triagular
network are composed of equilateral polygons, a fact that follows in
accordance with the $Z_3$ symmetry.

\begin{figure}
\centerline{\psfig{figure=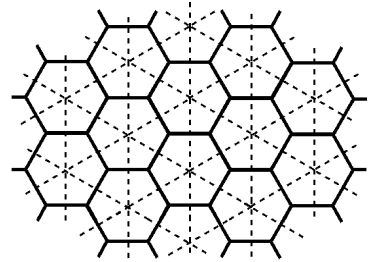,height=3.5cm}} 
\vspace{0.3cm}
\caption{The equilateral hexagonal network of defects, that may live
inside the host domain wall. The dashed lines show the dual lattice,
formed by equilateral triangles.}
\end{figure}

We now explore the breaking of the $Z_2\times Z_3$ symmetry
of the model. The simplest case refers to the breaking of the $Z_3$ symmetry,
without breaking the remaining $Z_2$ symmetry. We consider the case
of breaking the internal $Z_3$ symmetry in the following way. We take
for instance the vacuum state $v_1^{in}=(3/2)\,(1,0)$, and change
its position to a location farther from or closer to the other minima of
the system, increasing or decreasing the angle between two of the three
defects; see Fig.~5. We can do this with the inclusion in the
potential of another term, proportional to the second-order power on $\phi$.
We notice that the energy of the defect depends on the distance between the
two minima the defect connects, and goes with the cube
of it. Thus, if the vacuum state deviates significantly from its
$Z_3$-symmetric position, we cannot neglect the correction to the energy
of the defects. This changes the regular hexagonal pattern
of Fig.~4 to two other hexagonal patterns, composed of thicker
or thinner hexagons. We recall that hexagonal patterns may appear
in chemical systems \cite{osw91}, and in fluid convection \cite{abc93}
where they may also involve non-equilateral hexagons.

\begin{figure}
\centerline{\psfig{figure=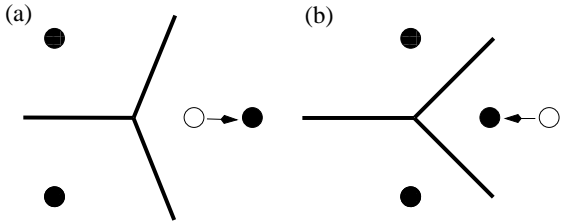,height=3.0cm}}
\vspace{0.2cm}
\caption{The vacuum states (black dots) and the junction that
forms the nested network. (a) and (b) illustrate the only two ways
of breaking the $Z_3\to Z_2$ symmetry.}
\end{figure}

We explore the presence of local defects in the hexagonal network
by introducing penta-hepta pair of cells, in a local deformation of
the network that disorganize its otherwise regular pattern. The mechanism is
similar to that of Refs.~{\cite{hne80,nel83}}. However, if the $Z_2$ symmetry
that governs the host domain wall is effective, local deformations may only
appear in a flat surface, requiring the pentagons and heptagons are not
regular polygons. This possibility may be seen in B\'enard-Marangoni
convection; see Ref.~{\cite{ogp83}} for a report on the experimental
observation of such paterns. But if together with the slight breaking
of the $Z_3$ symmetry of the internal network, one slightly breaks the $Z_2$
symmetry of the host domain wall locally, this will ultimately favor the
appearence of local deformations composed of pair of equilateral pentagons
and heptagons. Since in the network of equilateral hexagons, the presence of
equilateral pentagons and heptagons introduce local curvature, positive
and negative, respectively, we can understand these local defects as a
mechanism for roughening the planar surface that contains the network.
To break the symmetry of the nested network, one requires a slight change
of position of one of the three minima of the nested system, so we can
neglect the difference in energy and consider the tension as in the
regular hexagonal network. We see that the roughening springs to generate
higher energy states from the planar regular hexagonal structure.

We now concentrate on breaking the $Z_2$ symmetry of the host domain wall.
We can do this with the inclusion in the potential of a term odd in $\sigma$,
that slightly removes the degeneracy of the two minima $\sigma=\pm 3/2$.
Thus, the host domain wall bends trying to involve the local minimum, the
false vacuum. To stabilize the non-topological structure we include charged
fields into the system. The way one couples the charged fields is not unique,
but if we choose to add fermions, we can couple them to the $\sigma$ field
in a way such that the projection with $(\phi,\chi)\to(0,0)$ may leave the
model supersymmetric. This is obtained with the superpotential $W(\sigma)$,
with the Yukawa coupling $d^2W/d\sigma^2=(4/3)\sqrt{3}\,\sigma$. In this case
massless fermions bind \cite{jre76} to the host domain wall, and contribute to
stabilize \cite{mba96} the non-topological defect that emerges with the
breaking of the $Z_2$ symmetry.

The breaking of the $Z_2$ symmetry can be done breaking or not the remaining
$Z_3$ symmetry of the model. We examine these two possibilities supposing
that the host domain wall bends under the assumption of spherical symmetry,
becoming a non-topological defect with the standard spherical shape.
This is the minimal surface of genus zero, and according to the Euler
theorem we can only tile the spherical surface with three-junctions
as a regular polygonal network in the three different ways: with 4 triangles,
or 6 squares, or yet 12 pentagons. They are the tetrahedron, cube, and
dodecahedron, respectively. They are three of the five different
ways of tiling the sphere with regular polygons, known as the
Platonic solids \cite{poly}. These three cases preserve the $Z_3$
symmetry of the original network, locally, at the three-junction points.
However, if locally one slightly breaks the $Z_3$ symmetry of the
network to the $Z_2$ one, the three-junctions can now tile the spherical
surface with 12 pentagons and 20 hexagons. We think of breaking the $Z_3$
symmetry minimally, to the $Z_2$ symmetry, through the same mechanism
presented in Fig.~5. Thus, if the symmetry is broken slightly we can
consider the defect tensions as in the regular hexagonal network.

The tiling with 12 pentagons and 20 hexagons generates a spherical structure
that resembles the fullerene, the buckyball composed of sixty carbon
atoms \cite{fule,fuleb}. This is the truncated icosahedron, one among
thirteen different possibilities of tiling the sphere with regular polygons
of two or more distinct types, known as the Archimedean solids \cite{poly}.
The truncated icosahedron is one of the seven Archimedean solids constructed
with triple junctions, and it is the one that breaks the $Z_3$ symmetry
very slightly. 
We visualize the symmetries involved in the spherical structures thinking
of the corresponding dual lattices, which are triangular lattices, but in
the three first cases the triangles are equilateral, while in the fourth case
they are isosceles. We recall that regular heptagons introduce negative
curvature, so they cannot appear when the genus zero surface is minimal.
However, they may for instance spring to generate higher energy states from
the fullerene-like structure, locally roughening the otherwise smooth
spherical surface.

We write the energy of the non-topological structure as
$E^n_{nt}=E^s_{nt}+E_n$, where $E^s_{nt}$ stands for the energy of the
standard non-topological defect, and $E_n$ is the portion due to the nested
network.  We use $E^s_{nt}=E_q+E_h$, which shows the contributions of
the charged fields and of the host domain wall, respectively. We have
$E_h=S\,t_h$, and $E_n=N\,d\,t_{n}$, where $S$ is the area of the spherical
surface, and $N$ and $d$ are the number and length of segments in the nested
network. We introduce the ratio $E^n_{nt}/E^s_{nt}=1+[N/(1+r)](t_n/t_h)(d/S)$,
with $r=E_q/E_h$. The non-topological structure
nests a network of defects, which modifies the scenario one gets with the
standard domain wall. The modification depends on the way one couples charged
bosons and fermions to the $\sigma,\phi$, and $\chi$ fields. However, if the
$Z_3$ symmetry is locally broken to the $Z_2$ one, the most probable defect
corresponds to the fullerene or buckyball structure. But if the $Z_3$
symmetry is locally effective, there may be three equilateral structures,
the most probable arising as follows. We consider the simpler case
of plane polygonal structures, identifying the tetrahedron ($i=3$),
cube ($i=4$), and dodecahedron ($i=5$). We introduce $r_{ij}$ as the energy
ratio for the $i$ and $j$ structures. We get
$r_{ij}=(1+r+t_n/h_i\,t_h)/(1+r+t_n/h_j\,t_h)$, for $i,j=3,4,5$.
Here $h_3,h_4$, and $h_5$ stand for the radius of the
{\it incircle} of the triangle, square, and pentagon, respectively. Energy
favors the triangular lattice as the nested network. This configuration is
self-dual, because the network and its dual are the very same triangular
lattice. The two other configurations (that add to give the five Platonic
solids) are the octahedron, dual to the cube, and
the icosahedron, dual to the dodecahedron. They do not appear in
the $Z_2\times Z_3$ model because they require four- and five-junctions,
respectively.

Our work can be extended in several directions. For instance,
we could use the $Z_2\times Z_k$ symmetry $(k=4,5,6)$, getting to
$k$-junctions. This allows to tile the plane with squares $(k=4)$,
or triangles $(k=6)$, and the spherical surface with triangles, as the
octahedron $(k=4)$ or the icosahedron $(k=5)$. This direction seems
appropriate to model the recent experimental observations of squares
in specific Rayleigh-B\'enard and B\'enard-Marangoni
convections \cite{bln98,tok00}.
Also, in the $Z_2\times Z_3$ model, if the host domain wall bends
cylindrically, one may get to nanotube-like configurations \cite{nano,sdd99}.
As one knows, in certain types of nanotubes one can find polarons \cite{ufrj},
so we are using this idea to investigate the presence of chiral polarons
in chiral nanotubes. Another line of investigations concern the use of real
scalar fields to map laser beams interacting inside nonlinear fiber optics,
as for instance in the model we have already presented in Ref.~{\cite{lbb}},
where laser beams that generate black solitons are used to give rise to a
vector soliton of the black and bright type.

We would like to thank C.A.G. Almeida, B. Baseia, F.A. Brito, W. Freire,
L. Losano, J.M.C. Malbouisson, J.R.S. Nascimento, R.F. Ribeiro, M.M. Santos,
and C. Wotzasek for invaluable discussions, and CAPES, CNPq, PROCAD and PRONEX
for financial support.

\end{document}